\documentclass[preprint,aps,prd,groupedaddress,showpacs]
              {revtex4-1}%

\usepackage{graphicx}
\usepackage{amsmath}
\usepackage{bm}
\usepackage{relsize}
\RequirePackage{xspace}
\begin{document}

\title{
Leptonic Decay of $\Upsilon$, a Possible Signature of New Physics}



\author{Yu-Jie Zhang $~^{(a)}$, Hua-Sheng Shao $~^{(b)}$}
\affiliation{ {\footnotesize (a)~Key Laboratory of Micro-nano
Measurement-Manipulation and Physics (Ministry of Education) and School of Physics, Beihang University,
Beijing 100191, China}\\{\footnotesize (b)~Department of
Physics and National Key Laboratory of Nuclear Physics and
Technology, Peking University,
 Beijing 100871, China}}




\begin{abstract}
We  calculate the inclusive decay width of $\Upsilon \to l^+ l^-$.
Then we get the ratio  $R_{\tau\mu}=\Gamma[\Upsilon \to
\tau^+\tau^-]/ \Gamma[\Upsilon \to \mu^+\mu^-]$  to
${\cal{O}}(\alpha)$ and ${\cal{O}}(\alpha_s^2)$
within the Standard Model(SM). Comparing with the recent Babar's
data $R_{\tau \mu}=1.005\pm0.013\pm 0.022$, we find that SM
prediction $R_{\tau\mu}$ is  not consistent with the experimental
data in the error bar. The discrepancy is about $1.25\sigma$. So
leptonic decay of $\Upsilon$ may be a possible signature of New
Physics(NP). We present a better approach to test the Standard
Model, $R_{\tau\mu}(E_{soft})=\left.\Gamma[\Upsilon \to
\tau^+\tau^-+X]/ \Gamma[\Upsilon \to
\mu^+\mu^-+X]\right|_{E_X<E_{soft}}$. After
resumming  the large  logarithms , we get $R_{\tau\mu}(E_{soft})$
at the precision level of $0.1\%$. 
It can be compared with experimental data more
precise. We also consider the impact of $R_{\tau \mu}(E_{soft})$ and $R_{\tau \mu}$ from light
Higgs $h$ and pseudoscalar Higgs $A_0$. \end{abstract}

\maketitle




\section{Introduction}

Although the Standard Model(SM) of particle physics describes the interactions of
elementary particles very successfully, it is believed that SM is
not the final theory and there should be New Physics (NP) beyond SM.
So the hunting of NP is one of the  hottest  topics for theorist and
experimentalist. The B factories gave a very clear channel to test
SM, just as $\Upsilon(3S)\to \Upsilon(1S) \pi^+ \pi^-$, $\Upsilon\to
l^+ l^-$ ($l=\tau,\mu$). Recent Babar measured the ratio
\cite{Guido:2009xg,delAmoSanchez:2010bt}
\begin{eqnarray}
\label{Eq:Rllex}
R_{\tau \mu}=\frac{Br[\Upsilon\to \tau^+ \tau^-]}{Br[\Upsilon\to \mu^+ \mu^-]}=1.005\pm0.013\pm 0.022,
\end{eqnarray}
where branch ratio $Br[\Upsilon\to \tau^+ \tau^-(\mu^+ \mu^-)]$ is corresponded to inclusive decay width.
The final states radiations(FSR) effects due to photon(s) and gluon(s)
are taken into account in MC generator
.  The
Leading Order(LO) SM prediction of $R_{\tau \mu}$ is 0.992
\cite{SanchisLozano:2003ha,SanchisLozano:2002pm}. It is consistent with experimental date.  Then Babar claimed
``No significant deviation of the ratio $R_{\tau \mu}$ from the
SM expectation is observed''.

Theoretically, the high order corrections of the ratio $R_{\tau
\mu}$ should be taken into account. The SM predictions should be
compared with experimental data beyond tree level. At the same time, $R_{\tau \mu}$
is sensitively on the coupling of $h(A_0)b \bar b$ and $h(A_0)l^+l^-$  within NP.
It is an excellent probe for the new Higgs interactions in some NP Model,
where the coupling of Higgs $b \bar b$ and Higgs $l^+l^-$ is enhanced \cite{Accomando:2006ga}.
Then we should calculate the ratio $R_{\tau \mu}$ and compare with the
experimental data to test SM or hunt NP.

There are some theoretical and experimental works related with leptonic decay of $\Upsilon$.
The  Quantum Chromodynamics(QCD) corrections of $\Upsilon \to l^+l^-$ have been calculated to
two-loop \cite{Beneke:1997jm}. We have calculated $\Upsilon$ decay to charm
jet\cite{Zhang:2008pr}. The leptonic decay of vector bosons has been
calculated to Next-to-Leading Order(NLO) in Quantum Electodynamics(QED) \cite{Horejsi:1981kz}.
The CLEO got the ratio $R_{\tau \mu}= 1.02\pm 0.02\pm 0.05$ in 2006
\cite{Besson:2006gj}. The MC simulation of $\Upsilon\to \l^+ \l^-$
has been studied, where large logarithms have been
resummed\cite{Hamilton:2006xz}. The pseudoscalar Higgs $A_0$ is
also introduced in those
processes\cite{Domingo:2009tb,SanchisLozano:2003ha,Fullana:2007uq,SanchisLozano:2002pm}.
Babar has searched for a light Higgs boson $A_0$ in the radiative
decay of $\Upsilon(nS)\to \gamma A_0$, $A_0 \to l^+ l^-$ for
$n=1,2,3$. They found  no evidence for such processes in the mass
range $0.212 GeV \leq M_{A0} \leq 9.3GeV$ and no narrow structure
with  $4.03 GeV \leq M_{ \tau^+ \tau^-} \leq 10.10 GeV$
\cite{Aubert:2009cp,Aubert:2009cka}.

In this paper, we calculate the inclusive decay width of
$\Upsilon\to \l^+ \l^-$.  Then we get the precise prediction within
SM.  We also consider the impact from light Higgs $h$ and
pseudoscalar Higgs $A_0$.

\section{SM prediction}
\begin{figure}
\begin{center}
\includegraphics[width=7.5cm]{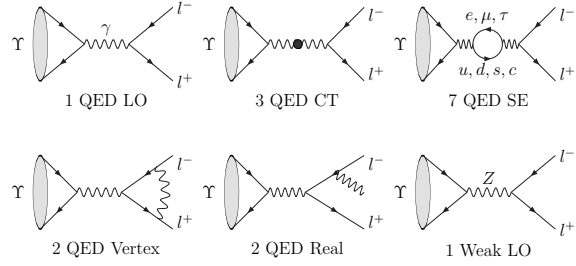}
\caption{\label{fig:bbllFeyn} Part of the Feynman diagrams of
$\Upsilon\to l^+l^-$ within Standard Model. }
\end{center}
\end{figure}

The LO QED Feynman diagrams of $\Upsilon \to l^+l^-$ are shown in
Fig.~\ref{fig:bbllFeyn}. Followed the process of $\Upsilon \to c
\bar c$ in Ref.\cite{Zhang:2008pr},  we can get the LO amplitude and
decay width of $\Upsilon \to l^+l^-$ ,
\begin{eqnarray}\label{eq:LOSOll}
{\cal{M}}_{LO}[\Upsilon \to l^+l^-]&=&\sqrt{\frac{16 \pi}{3M_\Upsilon^3}}\
 \alpha\ |R(0)|\   \bar l \not \! \epsilon \ l ,\nonumber \\
\Gamma_{LO}[\Upsilon \to l^+l^-]&=&\frac{4 |R(0)|^2 \alpha ^2 \sqrt{1-4 r_l} (1+2 r_l)}{9
   M_{\Upsilon }^2},
\end{eqnarray}
where $r_l= M_l^2/M_{\Upsilon }^2$, $|R(0)|$ is the radial wave
function of  $\Upsilon$ at origin, $\epsilon$ is the polarization
vector of $\Upsilon$. If expanded with $r_l$, we can get
\begin{eqnarray}\label{eq:LOSOGammall}
\Gamma_{LO}[\Upsilon \to l^+l^-]=\frac{4 |R(0)|^2 \alpha ^2 }{9
   M_{\Upsilon }^2}\left(1-6r_l^2+{\cal O}\left(r_l^3\right)\right).
\end{eqnarray}

We take into account the NLO QED correction here.
The renormalization of lepton and $b$ quark  wave function, and electron charge should appear.
We use $D=4-2\epsilon $ space-time dimension to regularize the divergence.
On-mass-shell (OS) scheme is selected for
$Z_{2b(l)}$ 
and modified minimal-subtraction
($~\overline{\rm MS}$~) scheme for $Z_e$:
\begin{eqnarray}\label{renDefrenccbb}
\delta Z_{2f}^{\rm OS}&=&-\frac{Q_f^2\alpha}{4\pi}
\left[\frac{1}{\epsilon_{\rm UV}}+\frac{2}{\epsilon_{\rm IR}}
-3\gamma_E+3\ln\frac{4\pi\mu^2}{M_f^2}+4\right],\nonumber \\
\delta Z_e ^{\overline{\rm MS}}&= &\frac{\alpha}{6 \pi}(3+\frac{10}{3})\left(\frac{1}{\epsilon_{\rm UV}}
-\gamma_E+\ln(4\pi)\right),
\end{eqnarray}
where $\mu$ is the renormalization scale, $\gamma_E$ is the Euler's
constant, $f=b,l$, and $Q_f$ is the charge of fermion $f$ in unit of electron charge.
The factor $3+\frac{10}{3}$ is from the charge and color factor of three flavor lepton $e, \mu, \tau$ ($3\times 1$)
and four flavor quark $u,d,s,c$ ($2\times 3 \times(1/9+4/9)$).
If we ignore the self energy of photon and the renormalization of
$\alpha$,  the NLO QED correction is just replaced $4\alpha_s/3$
with $\alpha$ from $\Upsilon \to c \bar c$\cite{Zhang:2008pr}.

For the corrections due to gluons in the final state are considered
in experimental Monte Carlo,  we should consider the QCD processes
$\Upsilon \to l^+l^-+gg$. We also consider
 NLO QCD corrections to the decay width of $\Upsilon \to l^+l^-$,
which give a factor of $1-4C_f\alpha_s/\pi$ to suppress the LO decay
width\cite{Beneke:1997jm}.

In numerical calculation, the parameters are selected as:
\begin{eqnarray}
M_e&=&0.5110MeV,\ \ M_d=0.00 MeV, \  \ M_u=0.00 MeV,   \nonumber \\
M_\mu&=&0.1057GeV,\ \ \ M_s=0.10GeV,\ \ \ M_c=1.30GeV,   \nonumber \\
M_\tau&=&1.7768GeV,\ \ \ M_b=4.73GeV,\ \ \ \alpha=1/132.33.
\end{eqnarray}
Here $ M_b=M_\Upsilon /2$. The renormalization scale $\mu$ is selected as $\mu=M_\Upsilon$,
and the fine structure constant $\alpha$  is calculated with the program alphaQED.f \cite{
Eidelman:1995ny}.
 The numerical $\Gamma[\tau (\mu)]$ and $R_{\tau \mu}$ are listed in
 Table.\ref{tab:SMinlcusive}. The LO prediction of $R_{\tau \mu}$ is
 0.992. It is used in Ref. \cite{Guido:2009xg,delAmoSanchez:2010bt},
  where claimed ``No significant deviation of the ratio $R_{\tau \mu}$ from the
SM expectation is observed''. But the QCD corrections should suppress the SM prediction
and drive $R_{\tau \mu}$ away from the experiment data.

\begin{table}\begin{center}
\caption{\label{tab:SMinlcusive}The numerical decay width of
inclusive processes  $\Upsilon \to l^+l^-$($l=\tau,\mu$) in unit of
$\frac{|R(0)|^2}{ 10^7 GeV^2}$ and $R_{\tau \mu}$  within SM.}
\begin{tabular}{r|ccc}
 &  $\Gamma[\tau]$ &
  $\Gamma[\mu]$ & $R_{\tau \mu}$\\\hline
LO &  2.822 & 2.844 & 0.992 \\
NLO QED& 2.777  & 2.798 &  0.993 \\
NLO QED, $l^+l^-gg$&2.780&2.836&0.980\\
NLO QED\&QCD, $l^+l^-gg$&1.743&1.791&$0.973\pm 0.001$\\
\hline Babar&-&-&$1.005\pm0.026$
\end{tabular}\end{center}
\end{table}

We should calculate the uncertainty for the theoretical prediction.
As an order estimate, one can get $R_{\tau \mu}^{LO} \sim {\cal
O}((\alpha/\pi)^0)$. For the NLO QED corrections have been taken into account, the uncertainty from higher order QED contributions is ${\cal O}(\alpha^2/\pi^2) \sim 6
\times 10^{-6}$
.
In the same way as QED,  the uncertainty from higher order QCD contributions is
${\cal O}(\alpha_s^3/\pi^3) \sim 3 \times 10^{-4}$
.
$Z$ can contribute to $\Upsilon \to l^+ l^-$ at tree level through
replacing photon with $Z$. 
We can get
\begin{eqnarray}
\label{eq:Zcontribution}
\frac{{\cal{M}}_{LO}^Z[\Upsilon \to l^+l^-]}{{\cal{M}}^\gamma_{LO}[\Upsilon \to l^+l^-]}=f_z\frac{
 \bar l\ [( 4\sin^2\theta_W-1 )\not \! \epsilon + \not \! \epsilon \gamma^5 ]\ l}{\bar l\ \not \! \epsilon \ l},
 \end{eqnarray}
 and
 \begin{eqnarray}
f_z=\frac{M_{\Upsilon }^2 \left(3-4 \sin ^2\theta
   _W\right)  }{16
   \left(M_{\Upsilon }^2-M_Z^2\right) \left(1-\sin
   ^2\theta _W\right) \sin ^2\theta _W}.
   \end{eqnarray}
Here $f_z \sim -M_{\Upsilon }^2/M_Z^2 \sim -10^{-2}$.
The vector current term should change the LO amplitude in Eq.\ref{eq:LOSOll} by a factor
$f_z \left(-1+4 \sin ^2\theta_W\right) \sim 10^{-3}$. It is just like replacing $\alpha$
with $\alpha (1-f_z (1-4 \sin ^2\theta_W))$ in Eq.\ref{eq:LOSOll},
but it is a global factor for three lepton at LO. Then the
uncertainty from $Z$  of $R_{\tau \mu}$ should be  ${\cal O}(f_z
\left(1-4 \sin ^2\theta_W\right) (R^{QED}_{\tau \mu}-R_{\tau
\mu}^{LO})) \sim {\cal O}( 10^{-6})$. Here superscript $QED$ means
NLO QED has been taken into account. The axial vector current is not
coherent with the vector current in Eq.\ref{eq:LOSOll}. It change
the width with a factor ${\cal O}(M_{\Upsilon }^4/M_Z^4)\sim {\cal
O}(10^{-4})$ and the ratio with a factor ${\cal O}(M_{\Upsilon }^2
M_l^2/M_Z^4)\sim {\cal O}(10^{-5})$ only. Compared with the $Z$
contribution at tree level, the other contributions from weak bosons
should be suppressed by $\alpha/\pi$ or more. Then the weak
contributions from $W^\pm, Z, H$ can be ignored safely. Within SM,
it should be considered that
 $\Upsilon \to \gamma \eta_b$,
where $\eta_b \to l^+l^-$ is followed \cite{Fullana:2007uq}. The
energy of $\gamma$  is about $70$ MeV in  $\Upsilon \to \gamma
\eta_b$ and $Br[\eta_b \to l^+l^-(+\gamma_{soft})] \sim
10^{-8}$\cite{Zhang:2009XX,Jia:2009ip}. For $\Upsilon \to \gamma
\eta_b$ is a P wave process, we can estimate  $Br[\Upsilon \to
\gamma \eta_b] $ through
\begin{eqnarray}
\frac{\Gamma[\Upsilon \to \gamma \eta_b]}{\Gamma[J/\psi \to \gamma \eta_c]} \sim \left(\frac{e_b}{e_c}\right)^2
\left(\frac{M_{J/\psi}(M_{\Upsilon}-M_{\eta_b})}{M_\Upsilon(M_{J/\psi}-M_{\eta_c})}\right)^3.
\end{eqnarray}
Then 
 $Br[\Upsilon \to \gamma \eta_b] \sim 10^{-5}$. So $Br[\Upsilon \to \gamma \eta_b]
 \times Br[\eta_b \to l^+l^-(+\gamma_{soft})]\sim 10^{-12}$. This can be ignored safely.
%
The uncertainties of $R_{\tau \mu}$  within SM are listed in
Tab.\ref{tab:SMuncertainties}. Then SM prediction is
\begin{eqnarray}\label{eq:SMRR}
R_{\tau \mu}=0.973\pm 0.001.
\end{eqnarray}
Compared with Eq.(\ref{Eq:Rllex}), it is  not consistent with the
experimental  data in the error bar. The discrepancy is about
$1.25\sigma$.

\begin{table}\begin{center}
\caption{\label{tab:SMuncertainties}The uncertainties of $R_{\tau
\mu}$  within SM.}
\begin{tabular}{r|cc}
 & Order &
  Numerical\\\hline
QED& $\alpha^2/\pi^2$ & $6 \times 10^{-6}$ \\
QCD& $\alpha_s^3/\pi^3$&$3 \times 10^{-4}$\\
$Z(W^\pm,  H)$ & $M_{\Upsilon }^2 M_l^2/M_Z^4$ or $\alpha M_l^2/(M_Z^2\pi)$& $4 \times 10^{-6}$\\
$\eta_b$&$Br[\Upsilon \to \gamma \eta_b]
 \times Br[\eta_b \to l^+l^-]$&$1 \times 10^{-12}$\\
\hline Total &-&$0.001$\\  \hline  \hline
$R_{\tau \mu}^{SM}$ &  1 & $0.973\pm 0.001$ \\
\end{tabular}\end{center}
\end{table}

\begin{figure}
\begin{center}
\includegraphics[width=7.5cm]{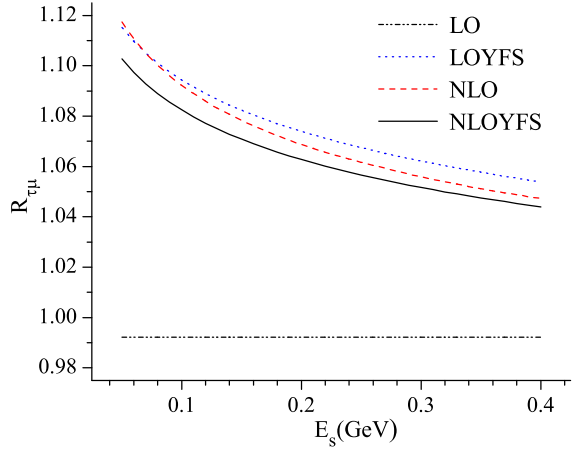}
\caption{\label{fig:Rtaomu}The dependence of $R_{\tau\mu}(E_{soft})$ on the soft cut $E_s$ within SM.}
\end{center}
\end{figure}

The QCD contributions have been taken into account in
Eq(\ref{eq:SMRR}). It is difficult to measure. So we present a
better approach to test the Standard Model,
$R_{\tau\mu}(E_{soft})=\left.\Gamma[\Upsilon \to \tau^+\tau^-+X]/
\Gamma[\Upsilon \to
\mu^+\mu^-+X]\right|_{M_X<E_{soft}}$. If we select
$E_{soft}\sim 5 GeV$, $\Gamma[\Upsilon \to
l^+l^-+gg]|_{M_X<E_{soft}}$ is less than $\Gamma[\Upsilon \to
l^+l^-]/1000$, then the impact on $R_{\tau\mu}(E_{soft})$ is less
than $2 \times 10^{-5}$, but the large logarithms appear
\begin{eqnarray}\label{Eq:largeL}
L= \ln \frac{4 E^2_{s}}{M^2_\Upsilon}\ln \frac{4 M^2_l}{M^2_\Upsilon}.
\end{eqnarray}
We resum the large logarithms with YFS resummation scheme\cite{Yennie:1961ad,Hamilton:2006xz},
\begin{eqnarray}
Y=\frac{-\alpha}{\pi } \left(2 \left(\ln
   r_l+1\right) \ln \frac{2
   E_{s}}{M_{\Upsilon }}+\frac{\ln
   {r_l}}{2} -\frac{\pi ^2}{3}+1\right).
\end{eqnarray}
The resumed results are
\begin{eqnarray}
\Gamma_{LO}^{res}&=&e^{Y}\ \Gamma_{LO},\nonumber \\
\Gamma_{NLO}^{res}&=&\left(e^{Y}-1-Y\right)\Gamma_{LO}+\Gamma_{QED}.
\end{eqnarray}
After the large  logarithms are resummed, we get $R_{\tau\mu}(E_{soft})$ with a soft cut at the
precision level of $0.1\%$.
The numerical $\Gamma_{\tau (\mu)}(E_{soft})$   in unit of $\frac{|R(0)|^2}{ 10^7 GeV^2}$
and $R_{\tau \mu}(E_{soft})$ with different energy cut $E_s$ are listed in Table.\ref{tab:SMGammaLL}.
The dependence of $R_{\tau\mu}(E_{soft})$ on the soft cut
$E_s$ is shown in Fig.\ref{fig:Rtaomu}. If we select $E_s=0.2GeV$. Including the uncertainty, the ratio is
\begin{eqnarray}\label{Eq:RllExclusive}
R_{\tau \mu}(0.2GeV)=1.0628 \pm 0.0011.
\end{eqnarray}
The effect of QCD is very weak in this channel. $R_{\tau\mu}(E_{soft})$
can be compared with experimental data more precise.

\begin{table}\begin{center}
\caption{\label{tab:SMGammaLL}The numerical decay width of processes
$\Upsilon \to l^+l^-$($l=\tau,\mu$) in unit of
$\frac{|R(0)|^2}{ 10^7 GeV^2}$ and $R_{\tau \mu}(E_{soft})$
within SM. $E_s=0.1$ means the soft cut is $0.1GeV$. }
\begin{tabular}{r|ccc}
 &  $\Gamma[\tau]$ &
  $\Gamma[\mu]$ & $R_{\tau \mu}(E_{soft})$\\\hline
LO &  2.8221 & 2.8444 & 0.9922 \\
LOYFS$|_{E_{s}=0.10}$ & 2.7277  & 2.4925  & 1.0944  \\
NLO$|_{E_{s}=0.05}$ & 2.6744  &  2.3932 &  1.1174 \\
NLOYFS$|_{E_{s}=0.05}$ & 2.6768  & 2.4272 & 1.1028 \\
NLO$|_{E_{s}=0.10}$ & 2.6954  & 2.4678  & 1.0922 \\
NLOYFS$|_{E_{s}=0.10}$ & 2.6970  & 2.4916  & 1.0824  \\
NLO$|_{E_{s}=0.20}$ & 2.7158  & 2.5411 & 1.0688 \\
NLOYFS$|_{E_{s}=0.20}$ & 2.7168  & 2.5564 &1.0628  \\
NLO$|_{E_{s}=0.45}$ & 2.7385  & 2.6236 & 1.0438 \\
NLOYFS$|_{E_{s}=0.45}$ & 2.7389  & 2.6312 &1.0409  \\
\end{tabular}\end{center}
\end{table}

%

\section{Impact from NP}
%
\begin{figure}
\begin{center}
\includegraphics[width=7.5cm]{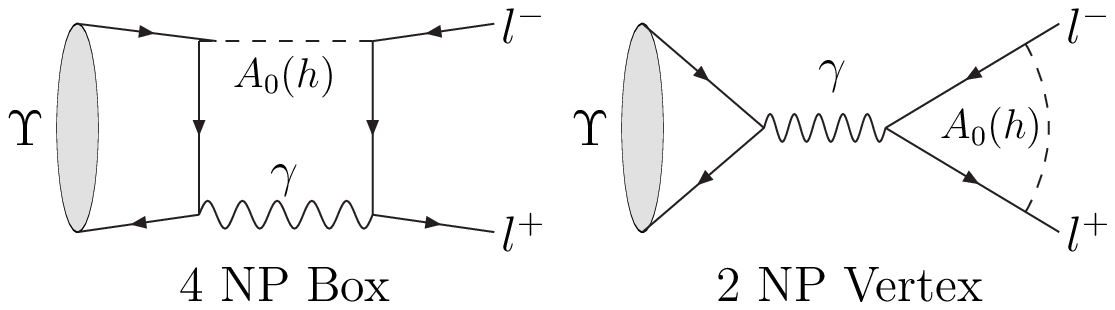}
\caption{\label{fig:bbllFeynA0hh} Part of the Feynman diagrams of
$\Upsilon\to l^+l^-$ which $A_0(h)$ involved.
The Feynman diagrams which exchange $A_0(h)$ between $b \bar b$ are
ignored for it should not change the ratio $R_{\tau\mu}$.}
\end{center}
\end{figure}

NP may play a role in the discrepancy between theoretical
prediction and experimental data of $R_{\tau\mu}$ in
Eq.(\ref{eq:SMRR}) and Eq.(\ref{Eq:Rllex}). We only consider the
scheme of light Higgs $h$ and pseudoscalar Higgs $A_0$ here.

\begin{figure}
\begin{center}
\includegraphics[width=7.5cm]{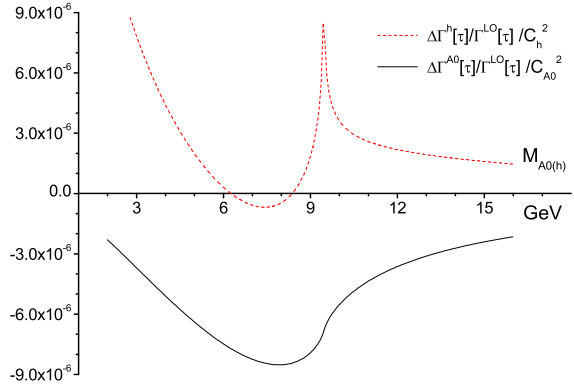}
\caption{\label{fig:GammaHiggs} The ${A_0(h)}$ impact on $\Upsilon\to \tau^+\tau^-$ as
a function of $M_{A0(h)}$.  The ${A_0(h)}$ impact on real contributions
ignored for it is suppressed by $E_s/M_b$ and $\Upsilon\to \mu^+\mu^-$
is ignored for it is suppressed by $M_\mu^2/M_\tau^2$. The Feynman diagrams which
exchange $A_0(h)$ between $b \bar b$ are ignored for it should not change the ratio $R_{\tau\mu}$.}
\end{center}
\end{figure}

The Feynman rules are $ i e M_f C_h /(2  M_W \sin \theta_W)\bar f f$
and $ - e M_f C_{A0} /(2  M_W \sin \theta_W)\bar f \gamma^5  f$ for
$h f \bar f$ and $A_0 f \bar f$ vertex respectively, here $f=l,b$.
$C_{A0(h)}$ are different in the special model, we consider them as
parameters. For it is IR finite which $A_0(h)$ involved in $\Upsilon
\to \gamma_{soft} \l^+\l^-$, so its contributions are suppressed by
$E_{s}/M_b\sim 4 \times 10^{-2}$ when compared with virtual
processes. So  we ignored the real processes and included the
virtual processes only  when we considered the impact of ${A_0(h)}$
to $R_{\tau\mu}(E_{soft})$.  The Feynman diagrams  are shown in
Fig.\ref{fig:bbllFeynA0hh}. The Feynman diagrams which exchange
$A_0(h)$ between $b \bar b$ are ignored for it should not change the
ratio $R_{\tau\mu}$.
  Compared with $\Gamma[\Upsilon \to \tau^+\tau^-]$,
the impact of ${A_0(h)}$ to $\Gamma[\Upsilon \to \mu^+\mu^-]$ is
suppressed by $M_\mu^2/M_\tau^2$, for the ${A_0(h)}l^+l^-$ coupling
and the spin flip between $\bar l \gamma^\nu l^-$ and  $\bar l
\gamma^5 l^-$ ($\bar l l^-$)  are both proportional to $M_l$. So  we
ignore the contributions for $\Upsilon \to \mu^+\mu^-$. Only
$\Upsilon \to  A_0^* \gamma^* \to \tau^+\tau^-$ and $\Upsilon \to
h^* \gamma^* \to \tau^+\tau^-$ are taken into account. The
contributions with  $A_0^* A_0^*$,  $h^* h^*$, or  $A_0^* h^*$ in
the loop are zero for $J^{PC}$. The numerical result of  the
${A_0(h)}$ impact as a function of $M_{A0(h)}$ from the loop Feynman
diagram is shown in Fig.\ref{fig:GammaHiggs}. The  ${A_0(h)}$ impact
on $R_{\tau\mu}$ is $R^{LO}_{\tau\mu}
\Gamma^{A0(h)}[\tau]/\Gamma^{LO}[\tau]$.

If we consider the $R_{\tau\mu}$, we should include the real
correction too.  
 If we select  $10.3GeV<M_{A0(h)}<10.6GeV$,
$\Gamma^{A0}[\tau]/\Gamma^{LO}[\tau] \sim  -4\times 10^{-6}C_{A0}^2
+ 5\times 10^{-10}C_{A0}^4$, and $\Gamma^{h}[\tau]/\Gamma^{LO}[\tau]
\sim  3\times 10^{-6}C_{h}^2 + 8\times 10^{-10}C_{h}^4$.

\section{Summary}
In summary, we calculate the inclusive decay width of $\Upsilon \to
l^+l^- $ ($l=\tau,\mu$). then we get the ratio
$R_{\tau\mu}=\Gamma[\Upsilon \to \tau^+\tau^-]/ \Gamma[\Upsilon \to
\mu^+\mu^-]$  to ${\cal{O}}(\alpha)$ and
${\cal{O}}(\alpha_s^2)$  within SM. Compared with the recent
Babar's data $R_{\tau \mu}=1.005\pm0.013\pm 0.022$, we find that SM
prediction $R_{\tau\mu}=0.973 \pm 0.001$ is not consistent with the
experimental data. The discrepancy is about $1.25\sigma$. So leptonic
decay of $\Upsilon$ may be a possible signature of NP.
 We present a better approach to test the Standard Model, $R_{\tau\mu}(E_{soft})=\left.\Gamma[\Upsilon \to \tau^+\tau^-+X]/
\Gamma[\Upsilon \to
\mu^+\mu^-+X]\right|_{E_X<E_{soft}}$. After
resumming the large  logarithms, we get $R_{\tau\mu}(E_{soft})$ with
a soft cut at the precision level of $0.1\%$. The effect of QCD is
very weak in this channel. It can be compared with experimental
data more precise. We also consider the possible solution, light
Higgs $h$ and pseudo scalar Higgs $A_0$. To clarify the discrepancy,
more work should be done by  theorist and experimentalist.

\begin{acknowledgments}
We thank Migue Angel Sanchis Lozano, Elisa Guido,
Chang-Zheng Yuan, Hai-Bo Li, Xin-Chun Tian, Zhao Li
for helpful assistance and discussions of the experiment.
This work was supported by
the National Natural Science Foundation of China (No 10805002).
\end{acknowledgments}




\providecommand{\href}[2]{#2}\begingroup\raggedright\endgroup

\end{document}